\documentstyle[12pt]{article}
\input epsf
\begin{document}
\begin{titlepage}

\centerline{\large \bf Quark Distribution in the Pion from}
\vspace{5mm}
\centerline{\large \bf QCD Sum Rules with Nonlocal Condensates}
\vspace{10mm}

\centerline{\bf A.V.Belitsky}
\vspace{10mm}

\centerline{\it Bogoliubov Laboratory of Theoretical Physics}
\vspace{3mm}
\centerline{\it Joint Institute for Nuclear Research}
\vspace{3mm}
\centerline{\it 141980, Dubna, Russia}
\vspace{20mm}

\centerline{\bf Abstract}

 We calculate the leading twist valence quark distribution in the
pion in the framework of QCD sum rules with nonlocal condensates.
Particular attention has been paid to the correct account for the
bilocal power corrections.
\end{titlepage}
%%%%%%%%%%%%%%%%%%%%%%%%%%%%%%%%%%%%%%%%%%%%%%%%%%%%%%%%%%%%%%%%%%%%%%%%%
\newpage

{\large \bf 1.} One of the most important achievements of Quantum
Chromodynamics is the determination of $Q^2$-evolution law for the
structure functions $F_i(x,Q^2)$ of deep inelastic scattering. It
allows one to calculate the magnitudes of observable $F_i(x,Q^2)$ at
some scale $Q^2$ starting from its value at another one $Q_0^2$.
The theoretical basis for this application is provided by factorization
theorems which give a possibility to express the physical cross
section as a "hard" parton subprocess convoluted with a "soft" parton
distribution function. While the former can be treated 
per\-tur\-ba\-ti\-ve\-ly
due to the celebrated asymptotical freedom, the latter is governed
by strong interaction dynamics at large distances and, therefore,
it cannot be evaluated within perturbative QCD. On the other hand,
the deep inelastic cross section can also be computed by using the
operator product expansion (OPE). This gives structure functions in
terms of certain coefficients multiplied by the target matrix elements
of local quark and gluon operators of definite twist. Combining
the two approaches allows one to express the parton distributions in
terms of quark and gluon correlation functions on the light cone.
Following Collins and Soper \cite{col82}, we can write for the
twist-2 valence quark distribution in a hadron
\begin{equation}
\langle h(p)|
{\cal O}\!\left(\frac{\lambda}{2} n,-\frac{\lambda}{2} n\right)
|h(p) \rangle
=4\int_{0}^{1}\cos(\lambda x)u_{h}(x),
\end{equation}
where\footnote{Throughout the paper $+$ subscript means the
convolution of the corresponding Lorentz index with light-cone
vector $n_{\mu}$, such that $n^2=0$, $(np)=1$, $(nq)=0$ and $q$
is $t$-channel momentum introduced below.}
\begin{equation}
{\cal O}\!\left(\frac{\lambda}{2} n,-\frac{\lambda}{2} n\right)\!=
\bar u\!\left(\frac{\lambda}{2} n\right)\!{\gamma}_+
\Phi\!\left[\frac{\lambda}{2} n,-\frac{\lambda}{2} n\right]\!
u\!\left(-\frac{\lambda}{2} n\right)\!
+(\lambda\rightarrow -\lambda)
\label{oper}
\end{equation}
and $\Phi$ is a path ordered exponential in the fundamental representation
of the colour group along the straight line which insures the gauge
invariance of the parton distribution
\begin{equation}
\Phi[{\rm x},{\rm y}]
=P\exp \left(ig({\rm x}-{\rm y})_{\mu}\int_{0}^{1}\!\!d\sigma
t^aB^a_{\mu}({\rm y}+\sigma ({\rm x}-{\rm y}))\right).
\end{equation}
It should be noted that the light-cone position representation is useful
to make contact with the OPE approach while the light-cone fraction
representation is appropriate for establishing the parton language.

The determination of parton distributions is, up to now, reserved
to experimental studies but as a final goal they are expected
to be evaluated from the first principles of the theory. In the
lack of complete understanding of the yet unclear confinement
mechanism they provide a challenging task for nonperturbative methods
presently available. Among the approaches which account for
nonperturbative effects the most close to QCD perturbation theory
are the QCD sum rules \cite{shi79}. In the last decade they were
applied with moderate success to determine nucleon and
photon structure functions in the region of intermediate values
of the Bjorken variable \cite{str84,str89} and, recently, in the small
$\lambda$-region for the light-cone position representation
\cite{bra95}.

While the nucleon structure functions are now well defined by the
analyses of the precise experimental data and are attacked
theoretically, much less is known about the parton distribution of
other hadrons, in particular of $\pi$-meson. Being of interest in
their own right they provide good testing ground for predictions
of the QCD sum rule method which will be used in the following
for the determination of the leading twist pionic valence quark
distribution.

{\large \bf 2.} In order to evaluate the quark distribution in the
pion by means of the QCD sum rules method, we consider an appropriate
three-point correlation function of two axial currents that have
non-zero projection onto the pion state being proportional to the
pion decay constant $\langle 0|j_{\mu}^{5}|\pi (p)\rangle
=if_{\pi}p_{\mu}$ and the nonlocal string operator $\cal O$ on
the light cone defined by eq. (\ref{oper})
\begin{equation}
W_{\mu\nu}(p_1,p_2,q)=i^2\int d^4{\rm x}d^4{\rm y}
e^{ip_1{\rm x}+iq{\rm y}}
\langle 0|
T\left\{
j_{\mu}^{5}({\rm x}),
{\cal O}\!\left({\rm y}+\frac{\lambda}{2} n,
{\rm y}-\frac{\lambda}{2} n\right)\!,
{j_{\nu}^{5}}^{\dagger} (0)
\right\}
|0 \rangle .
\label{corfun}
\end{equation}
The usual strategy is to use the duality between the hadronic and
partonic representation for the correlator under investigation.

On the one hand, we should consider the dispersion relation for
the latter and extract the contribution due to the low lying hadron,
namely, due to $\pi$-meson, approximating the higher
state contribution by perturbative spectral density
\begin{eqnarray}
&&W_{++}(p_1^2,p_2^2,Q^2=0)
=\frac{4f^2_{\pi}}
{(p_1^2-m^2_{\pi})(p_2^2-m^2_{\pi})}
\int_{0}^{1}dx \cos(\lambda x)u_{\pi}(x)\nonumber\\
&&+\frac{1}{{\pi}^2}\int_{0}^{\infty }\!\!\!\int_{0}^{\infty }\!\!
ds_{1}ds_{2}\frac{{\rho}_{pert}
(s_{1},s_{2},\lambda)}{(s_{1}-p_1^2)(s_{2}-p_2^2)}
(1-\theta (s_0-s_1)\theta (s_0-s_2)),
\end{eqnarray}
with parameter $s_0$ characterizing the beginning of the continuum.
Note that projecting the Lorentz indices of the pion interpolating
fields on the direction picked by vector $n_{\mu}$, we extract the leading
tensor structure in the infinite momentum frame. We omit the subtraction
polynomials in $p^2_1$ and $p^2_2$ because they disappear after the Borel
procedure has been applied. The latter leads to exponential suppression
of the excited state contribution in the phenomenological side of the
sum rule and gives factorial improvement of the OPE series at the
theoretical one. We perform the double Borel transformation and put
the parameters equal $M^2_1=M^2_2=2M^2$ in order not to introduce the
asymmetry between the initial and final pion states and to make contact
with two-point sum rules for the pion decay constant.

On the other hand, we consider the OPE for the same quantity.
Of course, the QCD sum rules with local condensates
are inappropriate here because the usual local power corrections
produce $\delta$-type contribution to the distribution function.
It is not surprising since some propagators are substituted by constant
factors that do not allow the momentum to flow and the whole hadron
momentum be carried by a single
quark. The probability density of this configuration in the phase
space is $\delta(1-x)$. Higher condensates produce even more singular
terms. However, this singular contribution can be smeared over the whole
region of the momentum fraction from zero to unity by avoiding the Taylor
expansion of the generic nonlocal objects which are the starting point
of all QCD sum rule calculations and introducing the concept of nonlocal
condensate \cite{shu82,mik89} which assumes the finite correlation
length for the vacuum fluctuations.

At the two-loop level, to which we restrict our analysis, we need
the bilocal quark and gluon condensates, trilocal quark-gluon
condensates and four-quark condensates. The latter will be factorized
into the product of bilocal scalar quark condensates via the vacuum
dominance hypothesis. For explicit calculations, it is convenient
to parametrize the bilocal condensates in the form of the well-known
$\alpha$-representation for propagators \cite{mik89}
\begin{eqnarray}
&&\langle 0|\bar {\psi}(0)\Phi[0,{\rm x}] \psi({\rm x})|0 \rangle
= \langle\bar {\psi}\psi\rangle
\int_{0}^{\infty} d\alpha f_S(\alpha) e^{\alpha {\rm x}^2/4}, \nonumber\\
&&\langle 0|\bar {\psi}(0)
\Phi[0,{\rm x}]\gamma_\mu \psi({\rm x})|0 \rangle =
-i{\rm x}_\mu
\frac{2}{81} \pi {\alpha}_s {\langle\bar {\psi}\psi\rangle}^2
\int_{0}^{\infty} d\alpha f_V(\alpha) e^{\alpha {\rm x}^2/4}.
\label{nlcon}
\end{eqnarray}

One comment concerning eqs. (\ref{nlcon}) is that in deriving a QCD sum
rule one can always perform a Wick rotation ${\rm x}_0
\rightarrow i{\rm x}_0$
and treat all the coordinates as Euclidean, ${\rm x}^2<0$. We use the
following ansatz for the distribution of vacuum quarks in the virtuality
$\alpha$ \cite{bak95}

\begin{equation}
f_S(\alpha)
=\frac{\sqrt{\gamma}}{2\Lambda K_1(2\Lambda\sqrt{\gamma})}
\exp\left(-\frac{{\Lambda}^2}{\alpha}-\alpha\gamma \right),
\end{equation}
which gives the exponential fall-off for the coordinate dependence
of the condensates found on a lattice \cite{gia84}. Here
$\Lambda^2=0.2GeV^2$,
and $\gamma$ is fixed from the lowest nontrivial moment of the
distribution function $f_S$ that is related to the value of the average
virtuality of the vacuum quarks $\lambda_q^2$ \cite{bel82}.

{\large \bf 3.} Conventional calculations of the perturbative diagram
(see fig. 1(a)) with the light quark masses neglected result in
\begin{equation}
{\Psi}_{pert} (M^2,Q^2,\lambda)=
\frac{3}{2{\pi}^2} M^2 \int_{0}^{1}dx \cos(\lambda x)
x \bar x \exp\left(-\frac{\bar x}{x}\frac{Q^2}{4M^2}\right).
\label{pert}
\end{equation}
Note that we have kept the $t$-channel momentum transferred to be
nonzero. If we expand the cosine in the Taylor series and integrate
over $x$, we find out that each moment possesses logarithmic
non-analyticities of the type ${\left(Q^2\right)}^n\ln Q^2$. These
terms come from the small-$x$ region, where the spectator quark
carries almost the whole momentum of the pion, so that the struck
quark becomes wee and can propagate over large distances in the
$t$-channel. Therefore, we have to perform additional factorization
for separation of small and large distances in the corresponding
invariant amplitude; this will lead to the appearance of additional terms
in the OPE for the three-point correlation function which correct the
small-$x$ dependence of the parton density.

The simplest nonperturbative correction comes from the vector
condensate (fig. 1(b))
\begin{equation}
{\Psi}_V (M^2,\lambda)=
\frac{8}{81}{\pi}{\alpha}_s{\langle\bar u u\rangle}^2
\int_{0}^{1}dx \cos(\lambda x)x f_V(\bar x M^2).
\end{equation}

The dominant contribution is due to the four-quark condensate.
For calculation of two loop diagrams appearing in the consideration
(fig. 1(c)) it is very convenient to use the following method which
is an extension of the calculation technique developed in ref.
\cite{mik88} for two-point correlators. The main ingredient is a
construction of a more general object, namely, the current in the vertex
opposite to the gluon propagator should be replaced by the nonlocal one
with light-like separation. The advantage of this substitution results
in appearing of extra $\delta$-function and introducing through this
replacement a set of variables which give the simplest integration.
At the end, we put the nonlocallity parameter equal zero. Performing
straightforward calculations we obtain
\begin{eqnarray}
{\Psi}_S (M^2,\lambda)\!\!&=\!\!&
\frac{32}{9}{\pi}{\alpha}_s{\langle\bar u u\rangle}^2
\int_{0}^{1}dx\cos(\lambda x)\nonumber\\
&&\times\int_{0}^{1}\!\!dy\!
\int_{0}^{1}\!\!d\xi\!
\int_{\frac{1}{2}}^{1}\!\!d\zeta
f_S\!
\left(
\frac{\bar x}{\xi}M^2
\right)\!
f_S\!
\left(
\frac{y}{\bar{\zeta}}M^2
\right)\!
\theta\!
\left(
\frac{\xi-\zeta}{x-y}
\right)\!
\frac{x\bar y}{|\bar xy\bar{\xi}\zeta
-x\bar y\xi\bar{\zeta}|}.
\end{eqnarray}

The gluon as well as trilocal quark-gluon condensate contributions
are numerically much less important than the power correction we
accounted for; therefore, we neglect them in what follows.

It is well known that there exists a parton sum rule that
implies that the pion contains one $u$-quark. Summing the
calculated contributions and taking the formal limit $Q^2\to 0$
in the perturbative term we can convince ourselves, comparing
the result with the sum rule for the pion decay
constant\footnote{Of course, the comparison should be made
with the sum rule by accounting for nonlocal condensates given in
ref. \cite{mik89}.}, that the normalization condition is broken.
The reason for this has already been mentioned earlier and we
elaborate this point below.

{\large \bf 4.} Now we derive a Ward identity (WI) \cite{bra95}
and show that the parton sum rule should be exact in QCD.
Of course, from the fact that $\cal O$ is a point-splitted
vector current it follows that in the limit $\lambda\to 0$ the
correlator
(\ref{corfun}) is related to the derivative of the two-point
correlation function of two axial currents. However, a more general
WI (for arbitrary $\lambda$) will be useful in the following for
discussion of the condensate contribution omitted.

Noticing that we are interested in the limit $Q^2=0$, we choose
vectors $n_{\mu}$ and $q_{\mu}$ to be proportional. Then,
integrating by parts in eq. (\ref{corfun}) and using the equation for
the complete variation of the phase factor with respect to the smooth
variation of the path $\Gamma\rightarrow {\Gamma}'$ :
${\rm x}_{\mu}(\tau)\rightarrow {{\rm x}'}_{\mu}(\tau)
={\rm x}_{\mu}(\tau)+ \delta {\rm x}_{\mu}(\tau)$ \cite{path78}
\begin{eqnarray}
&&\!\!\!\!\!\delta{\Phi}_{\Gamma}[{\rm x},{\rm y}]
={\Phi}_{{\Gamma}'}[{\rm x}',{\rm y}']-{\Phi}_{\Gamma}[{\rm x},{\rm y}]
=igt^aB^a_{\mu}({\rm x})\delta {\rm x}_{\mu}(1)
{\Phi}_{\Gamma}[{\rm x},{\rm y}]
-ig{\Phi}_{\Gamma}[{\rm x},{\rm y}]t^aB^a_{\mu}({\rm y})
\delta {\rm x}_{\mu}(0)\nonumber\\
&&\qquad\qquad\qquad+ig\int_{0}^{1}\!\! d\tau
{\Phi}_{\Gamma}[{\rm x},{\rm x}(\tau)]t^aG^a_{\mu\nu}({\rm x}(\tau))
\delta {\rm x}_{\mu}(\tau)
\frac{d{\rm x}_{\nu}(\tau)}{d\tau}{\Phi}_{\Gamma}[{\rm x}(\tau),{\rm y}]
\end{eqnarray}
with ${\rm x}_{\mu}(1)={\rm x}_{\mu}$, ${\rm x}_{\mu}(0)={\rm y}_{\mu}$,
we obtain
\begin{eqnarray}
&&\!\!\!\!\!\!\!\!\!\!\!W_{\mu\nu}
(p_1,p_2,q)\nonumber\\
&&\!\!\!\!\!\!\!\!\!\!\!=i\int\! d^4{\rm x}
e^{ip_1{\rm x}}\frac{{\rm x}_+}{(q{\rm x})}
\Bigl[
e^{iq{\rm x}}\langle 0|
T\left\{
j_{\mu}^{5}({\rm x},{\rm x}-\lambda n), {j_{\nu}^{5}}^{\dagger} (0)
\right\}
|0 \rangle
-\langle 0|
T\left\{
j_{\mu}^{5}({\rm x}), {j_{\nu}^{5}}^{\dagger}(0,\lambda n)
\right\}
|0 \rangle
\Bigr] \nonumber\\
&&\!\!\!\!\!\!\!\!\!\!\!+
\frac{\lambda}{2}\int_{-1}^{1}\!\! d\tau\!\!
\int\! d^4{\rm x}d^4{\rm y}
e^{ip_1{\rm x}+iq{\rm y}}
\frac{{\rm x}_+}{(q{\rm x})}
\langle 0|
T\left\{
j_{\mu}^{5}({\rm x}),{\cal G}\!\left({\rm y}+\frac{\lambda}{2} n,
{\rm y}-\frac{\lambda}{2} n,\tau \right)\!,
{j_{\nu}^{5}}^{\dagger}(0)
\right\}
|0 \rangle
+(\lambda \rightarrow -\lambda).\nonumber\\
&&
\label{ward}
\end{eqnarray}
Here
\begin{eqnarray}
&&\!\!\!\!\!\!\!\!\!\!{\cal G}\!\left({\rm y}+\frac{\lambda}{2} n,
{\rm y}-\frac{\lambda}{2} n,\tau \right)\nonumber\\
&&\!\!\!\!\!\!\!\!\!\!=\!\bar u\!\left({\rm y}+\frac{\lambda}{2} n\right)\!
\Phi\!
\left[{\rm y}+\frac{\lambda}{2} n,{\rm y}+\tau\frac{\lambda}{2} n\right]
\!{\gamma}_{\rho}
gt^aG_{\rho +}^a\!\left({\rm y}+\tau\frac{\lambda}{2} n\right)\!
\Phi\!
\left[{\rm y}+\tau\frac{\lambda}{2} n,{\rm y}-\frac{\lambda}{2} n\right]
\!u\!\left({\rm y}-\frac{\lambda}{2} n\right)\nonumber\\
&&
\end{eqnarray}
and
\begin{eqnarray}
j_{\mu}^{5}({\rm x},{\rm x}-\lambda n)
=\bar d({\rm x}){\gamma}_{\mu}{\gamma}_{5}
\Phi[{\rm x},{\rm x}-\lambda n]u({\rm x}-\lambda n)
\end{eqnarray}
is a point-splitted operator which when sandwiched between the
pion state and that of the vacuum, and convoluted with the
light-like vector $n_{\mu}$ defines the leading twist-2 pion
wave function. From this WI it follows that the normalization
of $u$-quark distribution is exact in QCD, provided it is not
spoiled by continuum subtraction
\begin{equation}
\int_{0}^{1}dxu_{\pi}(x)=1.
\label{normalization}
\end{equation}

{\large \bf 5.} As we have seen, in the limit $Q^2\to 0$ the perturbative
term though finite contains the logarithmic non-analyticities at
this point. This is a typical example of the mass singularities
in the QCD sum rules framework \cite{nes84,bel93,rus95}. In order
to get rid of this perturbative behaviour and replace it by a physical
one, it is necessary to modify the original OPE. For the form factor
type problem a two-fold structure of the modified OPE has been realized in
refs. \cite{bal82,yun82} being of the following schematic form:
\begin{eqnarray}
W(p_1^2,p_2^2,q^2)
=\sum_{d}C^{(d)}(p_1^2,p_2^2,q^2)
\langle {\cal O}_d \rangle
+\sum_{i} \int d^4{\rm x} e^{ip_1{\rm x}}{\cal C}^{(i)}({\rm x})
{\cal W}^i(q,{\rm x},\lambda).
\label{modope}
\end{eqnarray}
An additional second term determines the contribution due to the
long-distance propagation of quarks in the $t$-channel. Here ${\cal W}^i$
are the two-point correlators
\begin{equation}
{\cal W}^i(q,{\rm x},\lambda)
=\int d^4{\rm y} e^{iq{\rm y}}\langle 0|T\left\{
{\cal O}_i({\rm x},0),
{\cal O}\left({\rm y}+\frac{\lambda}{2} n,
{\rm y}-\frac{\lambda}{2} n\right)
\right\} |0\rangle
\end{equation}
of the operator in question and some nonlocal string operator of a definite
twist \cite{rus95} which arises from the OPE of $T$-product of pion
interpolating fields
\begin{equation}
T\left\{
j_{\mu}^{5}({\rm x}), {j_{\nu}^{5}}^{\dagger} (0)
\right\}
=\sum_{i} {\cal C}^{(i)}({\rm x}){\cal O}_i({\rm x},0).
\end{equation}
The coefficients $C^{(d)}(p_1^2,p_2^2,q^2)$ in eq. (\ref{modope}) are
free from non-analyticities or singularities in $Q^2$ because they
are defined as the difference between the original diagram and its
factorized expression which is the perturbative analogue of the
corresponding bilocal correlator. The bilocals cannot be directly
calculated in perturbation theory but we can write down the dispersion
relation for them
\begin{equation}
{\cal W}^i(q,{\rm x},\lambda)=\frac{1}{\pi}\int_{0}^{\infty}ds
\frac{{\rho}_i(s,({\rm x}q),{\rm x^2},\lambda)}{s-q^2},\label{biloc}
\end{equation}
accepting the conventional spectral density model: "low-lying hadron
plus continuum". The parameters of the model could be found from
auxiliary sum rules. There is no need in additional
subtractions in eq. (\ref{biloc}) because one always deals with the
difference between the "exact" bilocal and its perturbative part; so due
to the coincidence of their UV behaviours the subtraction terms cancel in
this difference.

{\large \bf 6.} The simplest bilocal power correction is given
by the following convolution:
\begin{equation}
W^{(1)}_{BL}(p_1^2,p_2^2,Q^2,\lambda)
=\int\!\!d^4{\rm x} e^{ip_1{\rm x}}{\cal C}_+^{(1)}({\rm x})
{\cal W}_{++}^{V}(q,{\rm x},\lambda n),
\label{conv}
\end{equation}
where the coefficient function is expressed through the quark
propagator ${\cal C}^{(1)}({\rm x})=2S_+({\rm x})$ and
\begin{equation}
{\cal W}_{++}^{V}(q,{\rm x},\lambda n)
=i\int\!\!d^4{\rm y} e^{iq{\rm y}}
\langle 0|
T\left\{
\bar u(0){\gamma}_+\Phi[0,{\rm x}]u({\rm x}), \>
{\cal O}\left({\rm y}+\frac{\lambda}{2} n,{\rm y}-\frac{\lambda}{2} n\right)
\right\}
|0\rangle .
\end{equation}
We extract the contact term \cite{yun82,bdy82} due to the vector
condensate from this correlator and saturate the remaining
part\footnote{Although for the present problem the calculation of
this part is only of academic interest, as it vanishes in the forward
limit being proportional to $Q^2$, we nevertheless evaluate it in
order to demonstrate the difficulties one faces when the contact-type
contribution is absent and the estimation of the correlator is carried
out saturating it by contribution of physical states.}
by the contributions of the mesons of increasing spin; these are
${\rho}^0$, $g$ states and so on. The fact that we are
interested in the $C$-odd distribution (valence quark) results in
contribution of spin-odd states in the $t$-channel. It is very
convenient to parametrize the appearing matrix elements via the wave
functions describing the light-cone momentum fraction distribution
of quarks inside mesons. To the leading twist accuracy we can write
\begin{eqnarray}
\langle 0|
\bar{\psi}(0)\Phi[0,{\rm x}]{\gamma}_{\mu}\psi ({\rm x})
|M^J(q,\eta)\rangle
&=&{\epsilon}^{(\eta )}_{\mu{\mu}_1{\mu}_2...{\mu}_{J-1}}
{\rm x}_{{\mu}_1}{\rm x}_{{\mu}_2}...{\rm x}_{{\mu}_{J-1}}
\left(m_M\right)^Jf^{(1)}_J{\phi}^{(1)}_J({\rm x}q)\nonumber\\
&-&iq_{\mu} {\epsilon}^{(\eta )}_{{\mu}_1{\mu}_2...{\mu}_J}
{\rm x}_{{\mu}_1}{\rm x}_{{\mu}_2}...{\rm x}_{{\mu}_J}
\left(m_M\right)^Jf^{(2)}_J{\phi}^{(2)}_J({\rm x}q),
\end{eqnarray}
where $J$ is a spin of the meson, $\eta$ its polarization and
${\epsilon}^{(\eta )}_{\mu{\mu}_1{\mu}_2...{\mu}_{J-1}}$ is a
polarization tensor. Inspired by our knowledge that in some cases
the asymptotical wave functions turn out to be rather close to
"exact" ones, we take in our estimation the former in the following
forms that are governed by conformal arguments $(\bar\beta\equiv 1-\beta)$
\begin{eqnarray}
{\varphi}^{(1)}_J(\beta)=\frac{\Gamma(2J+2)}{\Gamma^2(J+1)}
\left({\beta}\bar{\beta}\right)^J,\hspace{0.5cm}
{\varphi}^{(2)}_J(\beta)=\frac{\Gamma(2J+4)}{\Gamma(J+1)\Gamma(J+2)}
\left({\beta}-\bar{\beta}\right)
\left({\beta}\bar{\beta}\right)^J.
\end{eqnarray}

 In our model for the bilocal correlator we can achieve
this result if we assume the duality of the
meson resonances to the bare quark loop. In general,
this quite severe assumption turns out to be reasonable
for the case at hand, at least for the mesons of the lowest
spins $J$. It is known experimentally that the
physical cross section averaged over the $\rho$-meson peak
coincides with the quark one. Local duality for the low
lying states is a nontrivial dynamical property and is
not realized in all channels \cite{shi79a}. For the problem at
hand it can be explained by the specific interaction of the
classical vector mesons with the quark and gluon condensates
\cite{shi79}. The power correction for them even at
$M^2\approx m_{\rho}^2$ does not exceed $10-20\%$ of the main
perturbative term. So, $\rho$ is predicted to be dual to the
quark loop with the duality interval about
$s_{\rho}\approx 2m_{\rho}^2$. However, the local duality for
the higher spin mesons can be broken \cite{shi82}.

The net result for the difference between the "exact" bilocal
and its perturbative part reads
\begin{eqnarray}
&&\!\!\!\!\!\!\!\!\!\!\!\!\!\!{\cal W}_{++}^{V}(q,{\rm x},\lambda n)
-{\cal W}_{++}^{V(pert)}(q,{\rm x},\lambda n)\nonumber\\
&&\!\!\!\!\!\!\!\!\!\!\!\!\!\!=i{\rm x}_+\int_{0}^{1}d\tau
e^{i\tau (q{\rm x})}
\left\{
\langle 0| \bar u\!\left(\frac{\lambda}{2} n\right)\! {\gamma}_+
\Phi\!\left[\frac{\lambda}{2} n,{\rm x}-\frac{\lambda}{2} n\right]\!
u\!\left({\rm x}-\frac{\lambda}{2} n\right)\!
|0 \rangle + (\lambda\rightarrow -\lambda)
\right\}\nonumber\\
&&\!\!\!\!\!\!\!\!\!\!\!\!\!\!+2Q^2({\rm x}_+)^2\!\!
\int_{0}^{1}\!\!d\tau
\!\!\int_{0}^{1}\!\!d\beta\beta
e^{i\tau\beta (q{\rm x})}
\!\!\sum_{J=1,3,...}^{\infty}
\varphi^{(2)}_J(\beta)(-\lambda {\rm x}_+)^{J-1}\nonumber\\
&&\quad\quad\qquad\times\left\{
\frac{3}{8{\pi}^2}
\frac{\Gamma(J+2)}{2^{J-1}\Gamma(J)\Gamma(2J+4)}
\!\!\int_{0}^{\sigma_J^0}\!\!\!ds
\frac{s^J}{s+Q^2}+
(-1)^J\frac{(m_M^2)^Jf^{(1)}_Jf^{(2)}_J}{m_M^2+Q^2}
\right\},
\end{eqnarray}
where $\sigma_J^0$ is the continuum threshold and $m_M$ is the
mass of the lowest meson state in the channel of given spin $J$.
Substituting this expression into eq. (\ref{conv}) and performing
simple calculations we obtain (diagrammatical representation is given
in fig. 1(d))
\begin{eqnarray}
&&\!\!\!\!\!\!\!\!\!\!\!\!{\Psi}^{(1)}_{BL} (M^2,\lambda)=
\frac{8}{81}{\pi}{\alpha}_s{\langle\bar u u\rangle}^2
\int_{0}^{1}dx \cos(\lambda x)\bar x f_V(x M^2)
+Q^2e^{\frac{Q^2}{4M^2}}
\!\sum_{J=1,3,...}^{\infty}(i\lambda)^{J-1}\nonumber\\
&&\!\!\!\!\!\!\!\!\!\!\times\left\{
\frac{3}{8\pi^2}
\frac{1}{\Gamma(J)\Gamma(J+2)}
\!\!\int_{0}^{z^0_J}\!\!dz
\frac{z^J}{z+\frac{Q^2}{4M^2}}
+(-2)^J\left(\frac{m_M^2}{M^2}\right)^J\!\!
\frac{f^{(1)}_Jf^{(2)}_J}{m_M^2+Q^2}
\!\!\int_{0}^{1}d\beta
{\varphi}^{(2)}_J\left(\frac{1+\beta}{2}\right)
\right\},
\label{vect}
\end{eqnarray}
where $z^0_J=\sigma_J^0/4M^2$.
The former term is a contact-type contribution due to the vector
condensate. The first one in the curly brackets is the difference
between the perturbative analogue of the bilocal correlator and the
continuum contribution into the "exact" one. This part cancels the
logarithmic non-analyticities in the perturbative diagram
(eq. (\ref{pert})) corresponding to the leading twist-2 operator
in the OPE of pion currents. The tower of the next-to-leading
non-analyticities can be subtracted in a similar way by accounting
for the twist-4 operator. The last term displays the physical
contribution to the correlation function that possesses the correct
behaviour in the "momentum transferred $Q^2$". Requiring that in the
limit of large $Q^2$ the expression in the braces should be
zero, we come to the local duality relation for the overlaps
\begin{equation}
(m^2_M)^Jf^{(1)}_Jf^{(2)}_J=(-1)^{J-1}
\frac{3}{8\pi^2}\frac{J(\sigma^0_J)^{J+1}}{2^{J-1}\Gamma(2J+4)}.
\end{equation}

The sum of eqs. (\ref{pert}) and (\ref{vect}) is an analytical function
in $Q^2$ as all singularities are replaced by the combination
$Q^2+\sigma^0_J$ which is safe in the limit $Q^2\to 0$. Due to the
presence of the non-analyticities in each moment of the distribution
function, we need an infinite number of parameters to be found from
additional sum rules. Obviously, this is an impossible task. Safely,
for the problem at hand, this part vanishes in the forward limit and
the sum rule is dominated by the contact terms.

The result of eq. (\ref{vect}), as concerns the $Q^2$-independent part,
can be seen from the WI. The contact term contribution contained in the
bilocal correlator is effectively transformed into the power correction
due to vector condensate which arises together with eq. (\ref{pert}) from
the two-point correlation function in the WI (first two terms of eq.
(\ref{ward})). The latter was investigated in connection
with the pion wave function in the same framework \cite{mik89}. However,
it is not so for the most important bilocal part. In this respect the
WI is useless as it transforms the bilocals which can be reduced to the
condensates not accompanied by the strong coupling constant.

The dominant contribution comes from the bilocal correlator
convoluted with a three-pro\-pa\-ga\-tor coefficient function
(see fig. 1(e)) which looks like
\begin{equation}
W^{(2)}_{BL}(p_1^2,p_2^2,Q^2,\lambda)
=\int\!\!d^4{\rm x} e^{ip_1{\rm x}}
{\cal C}_+^{(2)}({\rm x}){\cal W}_+^{S}(q,{\rm x},\lambda n),
\end{equation}
(the mirror conjugated contribution can trivially be added)
where ${\cal C}_{\alpha}^{(2)}
=A{\rm x}_{\alpha}+Bn_{\alpha}+C{p_1}_{\alpha}$ and we
will not specify the coefficients in this decomposition because
of their complexity. All nonperturbative information is accounted
for in the correlator
\begin{equation}
{\cal W}_+^{S}(q,{\rm x},\lambda n)
=i\int\!\!d^4{\rm y} e^{iq{\rm y}}
\langle 0|
T\left\{
\bar u(0)\Phi[0,{\rm x}]u({\rm x}), \>
{\cal O}\!\left({\rm y}+\frac{\lambda}{2} n,
{\rm y}-\frac{\lambda}{2} n\right)\!\right\}
|0\rangle .
\end{equation}
In order to extract the contact term we make the following decomposition:
\begin{equation}
{\cal W}^{S}_{\mu}(q,{\rm x},\lambda n)
={\rm x}_{\mu}{\cal P}^S_{(1)}
+q_{\mu}{\cal P}^S_{(2)}
+\lambda n_{\mu}{\cal P}^S_{(3)}
\end{equation}
and convolute this expression with the vector $q_{\mu}$. Performing
the same steps as in the derivation of the Ward identity (\ref{ward})
we find
\begin{eqnarray}
&&\!\!\!\!\!\!\!\!\!\!\!\!\!\!\!\!\!\!\!\!\!\!(q{\rm x}){\cal P}^S_{(1)}
=i(q{\rm x})\int_{0}^{1}d\tau e^{i\tau (q{\rm x})}
\langle 0| \bar u\!\left(\frac{\lambda}{2}\right)\!
\Phi\!\left[\frac{\lambda}{2},{\rm x}-\frac{\lambda}{2}\right]\!
u\!\left({\rm x}-\frac{\lambda}{2}\right)\!|0 \rangle\nonumber\\
&&\!\!\!\!\!\!\!\!\!\!\!\!\!\!\!\!\!\!\!\!\!\!-i\frac{\lambda}{2}
\int_{-1}^{1}\!d\tau\!\int\! d^4{\rm y} e^{iq{\rm y}}
\langle 0|
T\left\{\bar u(0)\Phi[0,{\rm x}] u({\rm x}),
{\cal G}\!\left({\rm y}+\frac{\lambda}{2} n,
{\rm y}-\frac{\lambda}{2} n,\tau\right)\!\right\}
|0 \rangle
+(\lambda\rightarrow -\lambda)
+Q^2{\cal P}^S_{(2)}.
\end{eqnarray}
The last term in the second line vanishes in the limit of zero $Q^2$,
that manifests the absence of the massless particles in the corresponding
channels. Within the accuracy we are limited to, we are left with the first
term only because the second one contains an extra power of
$gG_{\mu\nu}$, and thus the corresponding OPE starts from the higher
orders in the coupling constant and the dimension of the operators.
Performing the integration by using the method outlined at the beginning
of the paper, we obtain the following contribution to the structure
function:
\begin{eqnarray}
{\Psi}^{(2)}_{BL} (M^2,\lambda)\!\!&=\!\!&
\frac{32}{9}{\pi}{\alpha}_s{\langle\bar u u\rangle}^2
\int_{0}^{1}dx\cos(\lambda x)\nonumber\\
&&\!\!\!\!\!\!\!\!\!\!\!\!\!\!\!\!\!\!\!\!\!\times\int_{0}^{1}\!\!dy\!
\int_{0}^{1}\!\!d\xi\!
\int_{0}^{\frac{1}{2}}\!\!d\zeta
f_S\!
\left(
\frac{\bar y}{\xi}M^2
\right)\!
f_S\!
\left(
\frac{x}{\bar{\zeta}}M^2
\right)\!
\theta\!
\left(
\frac{\zeta-\xi}{x-y}
\right)\!
\frac{y\bar x}{|\bar x y\xi\bar{\zeta}
-x\bar y\bar{\xi}\zeta|}.
\end{eqnarray}

Now, having accounted for additional terms in OPE, we can easily
check that the normalization condition for the quark distribution
in the pion is restored.

{\large \bf 7.} For zero $Q^2$ the perturbative spectral density
is concentrated on the line $s_1=s_2$, so that there is no
transition between the states with different masses.
We collect all contributions and make the continuum
subtraction that results in the substitution $M^2\to
M^2 (1-\exp(-s_0/M^2))$ in the perturbative term.
We have found good stability of the distribution
function with respect to the variation of the Borel parameter
in the region $0.5\leq M^2\leq 0.8$ for the standard value of
the continuum threshold $s_0=0.7GeV^2$. The normalization point
of the OPE is $\mu^2\sim 0.5 GeV^2$; therefore, the function
obtained can be regarded as an "input" quark distribution at
this low energy scale. In fig. 2, we present the curves
for the valence quark distribution in the pion for $M^2=0.6GeV^2$:
the solid and long-dashed lines correspond to the values of the
average virtuality of vacuum quarks
$\lambda_q^2=0.6GeV^2\ (\gamma^{-1}=0.154GeV^2)$ and
$\lambda_q^2=0.4GeV^2\ (\gamma^{-1}=0.087GeV^2)$,
respectively. In the large-$x$, region the corrections due to the
quark condensate do not exceed $30\%$ of the perturbative
term. However, in the small-$x$ region at $x=0.2$ the ratio of the
contact term to the main one comprises $50\%$ for $\lambda_q^2=0.6GeV^2$
and $70\%$ for $\lambda_q^2=0.4GeV^2$.
Below this point the nonperturbative contribution increases
and reaches $100\%$ at $x=0.13$ for $\lambda_q^2=0.4GeV^2$
(for $\lambda_q^2=0.6GeV^2$ it still amounts $50\%$). So,
for $x$ as small as $0.2$ we could not trust $x$-dependence of
our result. Of course, there is no possibility to reproduce
the correct $x\to 0$ behaviour of the parton density in the
present approach as it is determined by the exchanges of the
Regge trajectories.

Now we can comment on the contribution of the nonlocal gluon condensate
to our sum rule. As can be easily seen from the WI some part of
this contribution is concentrated in the two-point correlation
function, which has been studied in ref. \cite{mik93}. Being numerically
rather small, it contains terms not vanishing for $x\to 1$ as
distinguished from nonlocal quark condensates that do not spoil the
$(1-x)$-behaviour as $x\rightarrow 1$, but only renormalize the
slope. Therefore, the nonlocal gluon condensate limits the validity
of the present approach from the large-$x$ values. This conclusion is
made discarding additional terms appearing from the three-point
correlator in the WI which can somewhat change the situation. This
problem, as well as a particular value of $x$ in the large-$x$ region,
where the approach becomes invalid, deserves further investigation
and only smallness of the gluon condensate contribution favours our
decision to disregard it in the present study.

Since our result is valid only in the limited region of Bjorken
variable, we could not evolve it to the experimentally accessible
energies. In fig. 2, we compare our calculation with the distribution
obtained in the NJL model \cite{njl93} at the same normalization point
and find reasonable agreement between two approaches in a wide region
of the momentum fraction. In fig. 3, the latter evolved up to
$Q^2=20GeV^2$ (short-dashed curve) is compared with the presently
available fits of experimental data \cite{exper83}. It shows good
agreement with the result of the analysis of Sutton, Martin, Roberts
and Stirling (solid curve) \cite{newanal92}, which is consistent with
all present Drell-Yan and prompt photon $\pi N$ data. We also present
the (long-dashed) curve due to Gl\"uck, Reya and Vogt \cite{grv92};
however, their result does not agree with E615 experiment
\cite{exper83} which requires the valence distribution to be
larger by $20\%$. Similarly enhanced distribution has been obtained in
ref. \cite{cad90}. If the GRV curve is renormalized within a factor
of $1.2-1.3$ in the central region, there will be no disagreement between
the different analyses.

In conclusion, we have calculated the pionic parton density at low
momentum scale in QCD sum rules with nonlocal condensates.
It is shown that the parton sum rule is fulfilled only after
the bilocal power corrections are accounted for. We have
found good agreement with the $u$-quark distribution function
computed in the NJL model which when evolved up to the experimental
scales is well comparable with data.

\vspace{1cm}

We would like to thank S.V. Mikhailov and R. Ruskov for useful
discussions at an early stage of the work, A. Tkabladze
for help in numerical calculations and Prof. W.J. Stirling for
providing the Fortran package for the evolution of the pion
distribution extracted from the experimental data. This work
was supported by the Russian Foundation for Fundamental
Investigation, Grant $N$ 96-02-17631.

\newpage
\pagestyle{empty}

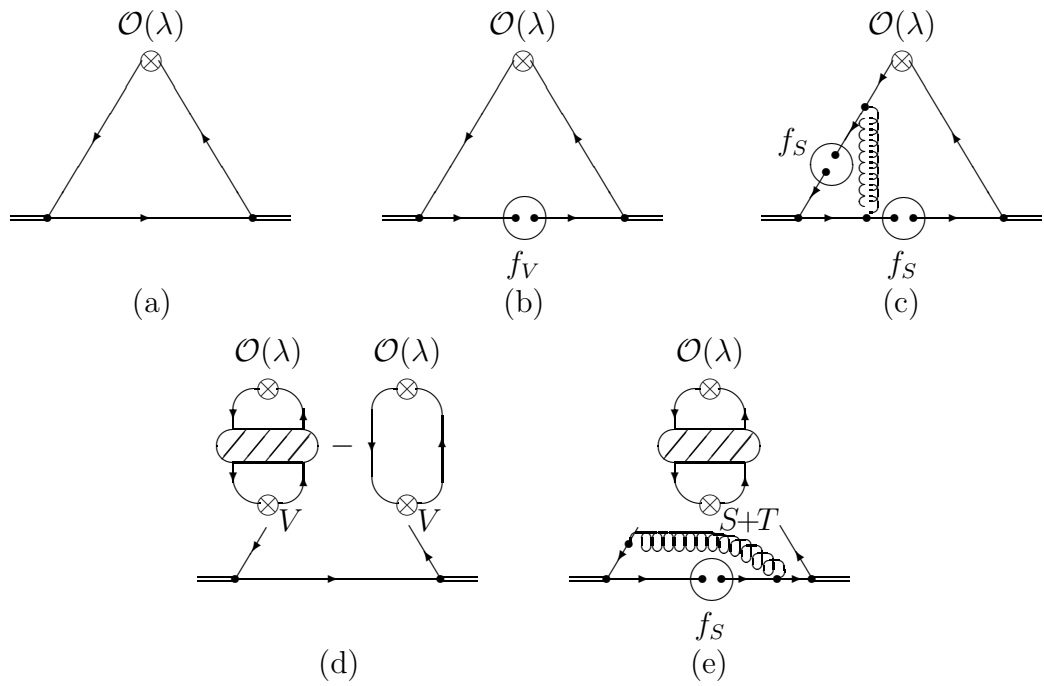
\begin{figure}[h]
\begin{center}
\unitlength=2.10pt
\special{em:linewidth 0.4pt}
\linethickness{0.4pt}
\begin{picture}(190.17,117.75)
\
\
\put(165.17,111.41){\makebox(0,0)[cc]{$\otimes$}}
\put(159.33,101.58){\oval(3.00,3.00)[r]}
\put(158.17,100.58){\oval(1.33,1.00)[l]}
\put(159.33,99.58){\oval(3.00,3.00)[r]}
\put(158.17,98.58){\oval(1.33,1.00)[l]}
\put(159.33,97.58){\oval(3.00,3.00)[r]}
\put(158.17,96.58){\oval(1.33,1.00)[l]}
\put(159.33,95.58){\oval(3.00,3.00)[r]}
\put(158.17,94.58){\oval(1.33,1.00)[l]}
\put(159.33,93.58){\oval(3.00,3.00)[r]}
\put(158.17,92.58){\oval(1.33,1.00)[l]}
\put(159.33,91.58){\oval(3.00,3.00)[r]}
\put(158.17,90.58){\oval(1.33,1.00)[l]}
\put(159.33,89.58){\oval(3.00,3.00)[r]}
\put(158.17,88.58){\oval(1.33,1.00)[l]}
\put(159.33,87.58){\oval(3.00,3.00)[r]}
\put(158.17,86.58){\oval(1.33,1.00)[l]}
\put(159.33,85.58){\oval(3.00,3.00)[r]}
\put(165.50,83.08){\circle{8.00}}
\put(152.50,92.75){\circle{8.00}}
\put(151.83,91.41){\vector(-2,-3){3.00}}
\put(151.50,91.41){\line(-3,-5){5.00}}
\put(163.50,111.41){\line(-3,-5){10.00}}
\put(166.50,111.41){\line(3,-5){17.00}}
\put(146.50,83.08){\line(1,0){17.33}}
\put(171.50,83.08){\vector(1,0){4.00}}
\put(149.50,83.08){\vector(1,0){3.67}}
\put(167.17,83.08){\line(1,0){16.33}}
\put(158.83,83.08){\circle*{1.33}}
\put(158.50,103.08){\circle*{1.33}}
\put(163.83,83.08){\circle*{1.33}}
\put(167.17,83.08){\circle*{1.33}}
\put(151.50,91.41){\circle*{1.33}}
\put(153.17,94.41){\circle*{1.33}}
\put(146.50,83.08){\circle*{1.33}}
\put(183.50,83.08){\circle*{1.33}}
\put(183.50,83.41){\line(1,0){6.67}}
\put(190.17,82.75){\line(-1,0){6.67}}
\put(146.50,83.41){\line(-1,0){6.67}}
\put(139.83,82.75){\line(1,0){6.67}}
\put(165.17,117.75){\makebox(0,0)[cc]{${\cal O}(\lambda)$}}
\put(96.92,111.41){\makebox(0,0)[cc]{$\otimes$}}
\put(97.25,83.08){\circle{8.00}}
\put(89.24,101.08){\vector(-2,-3){3.00}}
\put(98.25,111.41){\line(3,-5){17.00}}
\put(78.25,83.08){\line(1,0){17.33}}
\put(102.91,83.08){\vector(1,0){4.00}}
\put(82.58,83.08){\vector(1,0){3.67}}
\put(98.92,83.08){\line(1,0){16.33}}
\put(95.58,83.08){\circle*{1.33}}
\put(98.92,83.08){\circle*{1.33}}
\put(78.25,83.08){\circle*{1.33}}
\put(115.25,83.08){\circle*{1.33}}
\put(115.25,83.41){\line(1,0){6.67}}
\put(121.92,82.75){\line(-1,0){6.67}}
\put(78.25,83.41){\line(-1,0){6.67}}
\put(71.58,82.75){\line(1,0){6.67}}
\put(96.92,117.75){\makebox(0,0)[cc]{${\cal O}(\lambda)$}}
\put(29.92,111.41){\makebox(0,0)[cc]{$\otimes$}}
\put(31.25,111.41){\line(3,-5){17.00}}
\put(11.25,83.08){\circle*{1.33}}
\put(48.25,83.08){\circle*{1.33}}
\put(48.25,83.41){\line(1,0){6.67}}
\put(54.92,82.75){\line(-1,0){6.67}}
\put(11.25,83.41){\line(-1,0){6.67}}
\put(4.58,82.75){\line(1,0){6.67}}
\put(29.92,117.75){\makebox(0,0)[cc]{${\cal O}(\lambda)$}}
\put(95.25,111.41){\line(-3,-5){17.00}}
\put(22.25,101.08){\vector(-2,-3){3.00}}
\put(28.25,111.41){\line(-3,-5){17.00}}
\put(26.25,83.08){\vector(1,0){3.67}}
\put(11.25,83.08){\line(1,0){37.00}}
\put(96.91,74.58){\makebox(0,0)[cc]{$f_V$}}
\put(165.17,74.58){\makebox(0,0)[cc]{$f_S$}}
\put(145.50,96.75){\makebox(0,0)[cc]{$f_S$}}
\put(158.83,103.08){\vector(-2,-3){3.00}}
\put(163.83,111.74){\vector(-2,-3){3.00}}
\put(175.83,96.08){\vector(-2,3){1.67}}
\put(107.57,96.08){\vector(-2,3){1.67}}
\put(40.58,96.08){\vector(-2,3){1.67}}
\put(130.92,18.00){\circle{8.00}}
\put(116.91,25.67){\vector(-2,-3){3.00}}
\put(111.92,18.00){\line(1,0){17.33}}
\put(133.58,18.00){\vector(1,0){4.00}}
\put(115.92,18.00){\vector(1,0){3.67}}
\put(132.59,18.00){\line(1,0){16.33}}
\put(129.25,18.00){\circle*{1.33}}
\put(132.59,18.00){\circle*{1.33}}
\put(111.92,18.00){\circle*{1.33}}
\put(148.92,18.00){\circle*{1.33}}
\put(148.92,18.33){\line(1,0){6.67}}
\put(155.59,17.67){\line(-1,0){6.67}}
\put(111.92,18.33){\line(-1,0){6.67}}
\put(105.25,17.67){\line(1,0){6.67}}
\put(130.59,58.67){\makebox(0,0)[cc]{${\cal O}(\lambda)$}}
\put(130.58,9.50){\makebox(0,0)[cc]{$f_S$}}
\put(147.57,20.67){\vector(-2,3){1.67}}
\put(143.25,18.00){\vector(1,0){4.00}}
\put(118.08,24.56){\oval(3.67,3.67)[t]}
\put(120.08,24.56){\oval(3.67,3.67)[t]}
\put(121.08,24.39){\oval(1.67,2.67)[b]}
\put(122.08,24.56){\oval(3.67,3.67)[t]}
\put(119.08,24.39){\oval(1.67,2.67)[b]}
\put(123.08,24.39){\oval(1.67,2.67)[b]}
\put(124.08,24.56){\oval(3.67,3.67)[t]}
\put(125.08,24.39){\oval(1.67,2.67)[b]}
\put(126.08,24.56){\oval(3.67,3.67)[t]}
\put(128.08,24.56){\oval(3.67,3.67)[t]}
\put(129.08,24.39){\oval(1.67,2.67)[b]}
\put(130.08,24.56){\oval(3.67,3.67)[t]}
\put(127.08,24.39){\oval(1.67,2.67)[b]}
\put(131.08,24.39){\oval(1.67,2.67)[b]}
\put(132.08,24.23){\oval(3.67,3.67)[t]}
\put(134.08,23.89){\oval(3.67,3.67)[t]}
\put(133.08,24.06){\oval(1.67,2.67)[b]}
\put(136.08,23.23){\oval(3.67,3.67)[t]}
\put(135.08,23.39){\oval(1.67,2.67)[b]}
\put(138.08,22.22){\oval(3.67,3.67)[t]}
\put(137.08,22.40){\oval(1.67,2.67)[b]}
\put(137.92,22.33){\line(0,1){1.33}}
\put(140.08,20.89){\oval(3.67,3.67)[t]}
\put(139.08,21.07){\oval(1.67,2.67)[b]}
\put(139.92,20.66){\line(0,1){1.67}}
\put(115.92,24.33){\circle*{1.33}}
\put(117.58,27.33){\line(-3,-5){5.67}}
\put(143.25,27.67){\line(3,-5){5.67}}
\put(137.58,28.67){\makebox(0,0)[cc]{$S\!\!+\!\!T$}}
\put(51.00,52.33){\makebox(0,0)[cc]{$\otimes$}}
\put(50.99,27.34){\vector(-2,-3){3.00}}
\put(45.00,18.00){\circle*{1.33}}
\put(82.00,18.00){\circle*{1.33}}
\put(82.00,18.33){\line(1,0){6.67}}
\put(88.67,17.67){\line(-1,0){6.67}}
\put(45.00,18.33){\line(-1,0){6.67}}
\put(38.33,17.67){\line(1,0){6.67}}
\put(51.00,58.67){\makebox(0,0)[cc]{${\cal O}(\lambda)$}}
\put(80.65,20.67){\vector(-2,3){1.67}}
\put(50.66,27.33){\line(-3,-5){5.67}}
\put(76.33,27.67){\line(3,-5){5.67}}
\put(50.83,42.00){\oval(18.33,6.00)[]}
\put(51.00,31.33){\makebox(0,0)[cc]{$\otimes$}}
\put(52.50,45.00){\oval(9.67,14.67)[rt]}
\put(49.32,45.00){\oval(9.33,14.67)[lt]}
\put(52.83,39.00){\oval(9.00,14.67)[rb]}
\put(55.00,28.67){\makebox(0,0)[cc]{$V$}}
\put(60.00,18.00){\vector(1,0){3.67}}
\put(45.00,18.00){\line(1,0){37.00}}
\put(76.00,52.33){\makebox(0,0)[cc]{$\otimes$}}
\put(76.00,58.67){\makebox(0,0)[cc]{${\cal O}(\lambda)$}}
\put(76.00,31.33){\makebox(0,0)[cc]{$\otimes$}}
\put(77.50,45.00){\oval(9.67,14.67)[rt]}
\put(74.32,45.00){\oval(9.33,14.67)[lt]}
\put(77.83,39.00){\oval(9.00,14.67)[rb]}
\put(79.99,28.67){\makebox(0,0)[cc]{$V$}}
\put(69.65,41.67){\vector(0,-1){1.00}}
\put(82.33,42.68){\vector(0,1){1.00}}
\put(74.49,38.83){\oval(9.67,14.33)[lb]}
\put(49.50,38.83){\oval(9.67,14.33)[lb]}
\put(69.66,45.67){\line(0,-1){8.00}}
\put(82.32,37.67){\line(0,1){8.67}}
\put(44.66,47.67){\vector(0,-1){1.00}}
\put(44.66,36.34){\vector(0,-1){1.00}}
\put(57.34,47.68){\vector(0,1){1.00}}
\put(57.34,35.68){\vector(0,1){1.00}}
\put(64.33,42.00){\makebox(0,0)[cc]{$-$}}
\put(42.66,39.67){\line(5,6){4.33}}
\put(54.33,39.00){\line(5,6){4.33}}
\put(46.33,39.00){\line(5,6){5.00}}
\put(50.33,39.00){\line(5,6){5.00}}
\put(130.58,52.33){\makebox(0,0)[cc]{$\otimes$}}
\put(130.41,42.00){\oval(18.33,6.00)[]}
\put(130.58,31.33){\makebox(0,0)[cc]{$\otimes$}}
\put(132.08,45.00){\oval(9.67,14.67)[rt]}
\put(128.90,45.00){\oval(9.33,14.67)[lt]}
\put(132.41,39.00){\oval(9.00,14.67)[rb]}
\put(129.08,38.83){\oval(9.67,14.33)[lb]}
\put(124.24,47.67){\vector(0,-1){1.00}}
\put(124.24,36.34){\vector(0,-1){1.00}}
\put(136.92,47.68){\vector(0,1){1.00}}
\put(136.92,35.68){\vector(0,1){1.00}}
\put(122.24,39.67){\line(5,6){4.33}}
\put(133.91,39.00){\line(5,6){4.33}}
\put(125.91,39.00){\line(5,6){5.00}}
\put(129.91,39.00){\line(5,6){5.00}}
\put(135.92,22.67){\line(0,1){1.33}}
\put(141.08,20.07){\oval(1.67,2.67)[b]}
\put(142.00,19.67){\line(0,1){1.33}}
\put(142.08,19.56){\oval(3.67,3.67)[t]}
\put(142.67,19.66){\oval(2.67,2.67)[rb]}
\put(142.67,18.00){\circle*{1.33}}
\put(30.00,67.50){\makebox(0,0)[cc]{(a)}}
\put(97.08,67.50){\makebox(0,0)[cc]{(b)}}
\put(165.00,67.50){\makebox(0,0)[cc]{(c)}}
\put(63.75,2.50){\makebox(0,0)[cc]{(d)}}
\put(130.42,2.50){\makebox(0,0)[cc]{(e)}}
\end{picture}
\end{center}
\caption{Diagrams contributing to the operator product
expansion of the correlation function (4): the
first line display ordinary power corrections, while the
second one --- contribution due to the bilocal correlators.}
\end{figure}

\newpage

\begin{minipage}[t]{7.8cm} {
\begin{center}
\hspace{3cm}
\mbox{
\epsfysize=7.0cm
\epsffile[0 0 500 500]{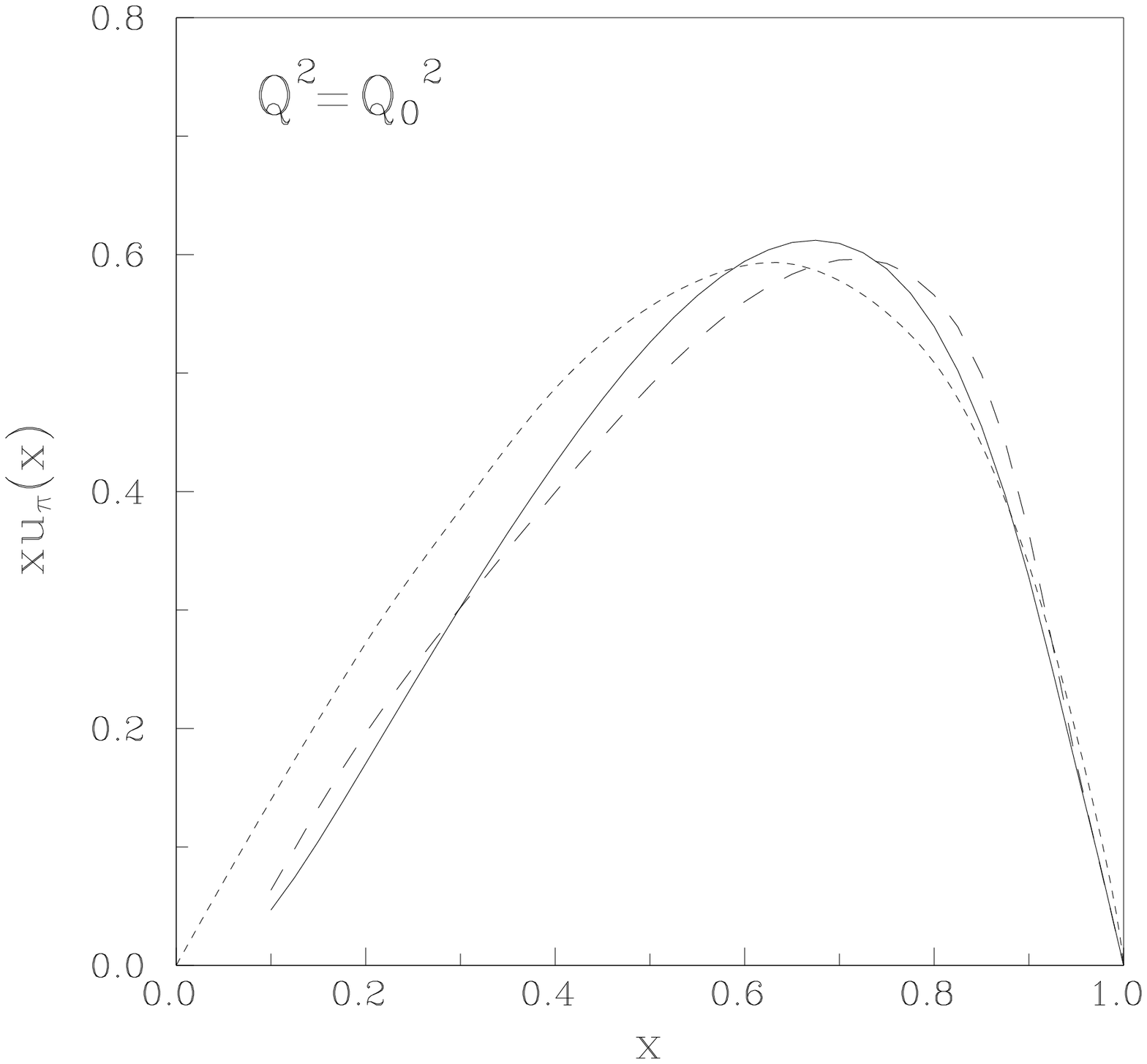}
}
\end{center}
}\end{minipage}

\vspace{0.5cm}

{\bf Figure.~2: } Quark distribution in the pion at the low energy scale
$\mu^2\sim 0.5GeV^2$ calculated from the QCD sum rule for different
values of the average virtuality of vacuum quarks: solid and
long-dashed curves correspond to $\lambda_q^2=0.6GeV^2$ and
$\lambda_q^2=0.4GeV^2$, respectively. Short-dashed curve is the
$u$-quark density found in the NJL model [22].

\vspace{1cm}

\begin{minipage}[t]{7.8cm} {
\begin{center}
\hspace{3cm}
\mbox{
\epsfysize=7.0cm
\epsffile[0 0 500 500]{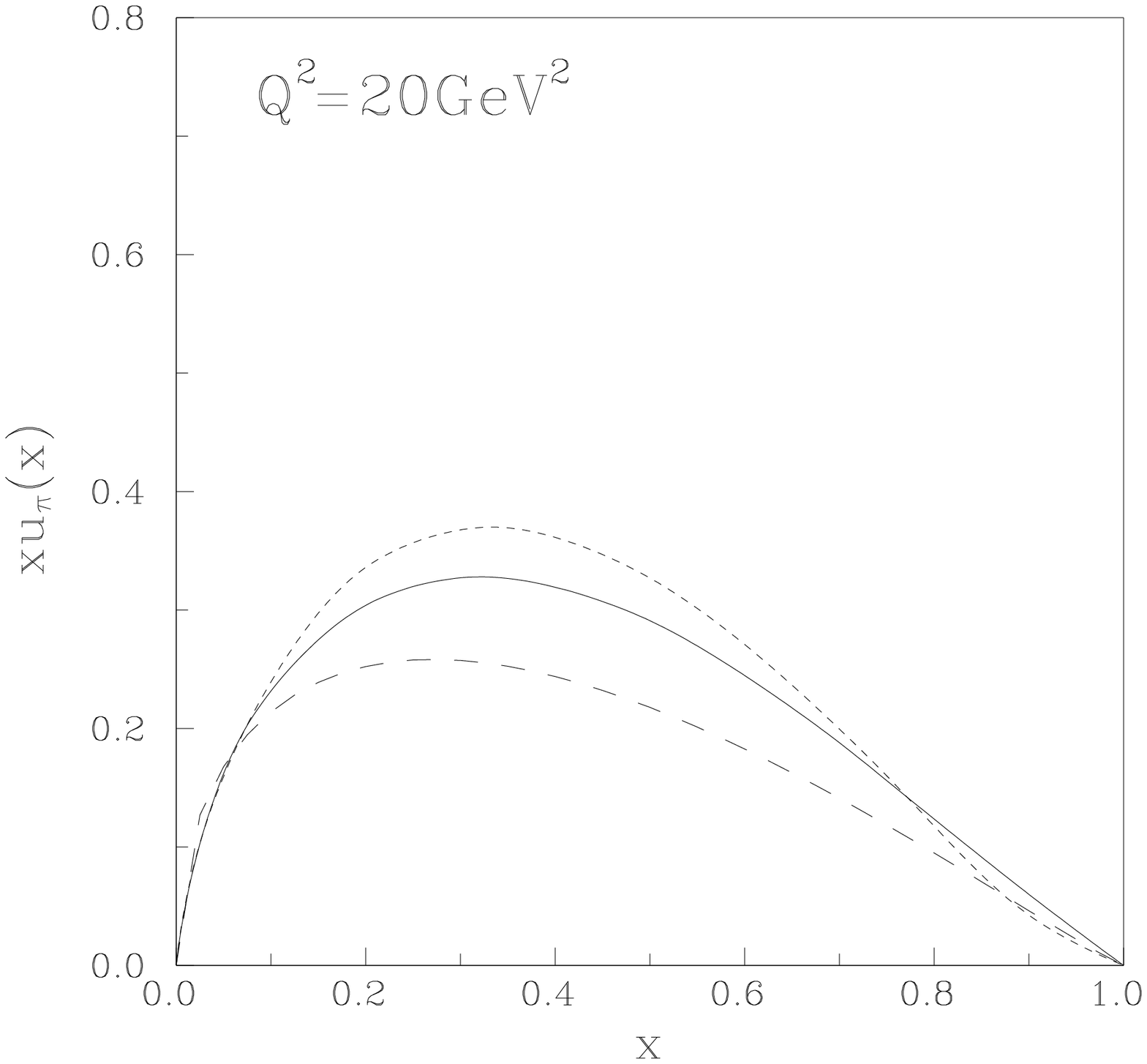}
}
\end{center}
}\end{minipage}

\vspace{0.5cm}

{\bf Figure.~3: }The experimental fits of the valence $u$-quark
distribution in the pion at $Q^2=20GeV^2$: SMRS result [24]
is depicted by solid curve, GRV analysis [25] is shown by
long-dashed one. Short-dashed curve is the $u$-quark density
calculated in the NJL model [22] evolved up to $Q^2=20GeV^2$.

\end{document}